# Scattering in the quenched approximation*

Claude Bernard and Maarten Golterman

Department of Physics, Washington University
St. Louis, MO 63130-4899, USA

We study, in the quenched approximation, Lüscher's relation between pion scattering lengths and the finite-volume energy of two pions at rest. The quenched relation is drastically different from the full theory one; in particular, "enhanced finite-volume corrections" of order $L^0 = 1$ and $L^{-2}$ occur at one loop ($L$ is the linear size of the box), due to the special properties of the $\eta'$ in the quenched approximation. Numerical examples show that the size of these effects can be substantial.

## 1. Introduction

Lüscher has shown that two pions at rest in a finite volume $L^3$ have an energy

$$E = 2m_\pi - \frac{4\pi a_0}{m_\pi L^3}\left(1 + c_1\frac{a_0}{L} + c_2\frac{a_0^2}{L^2}\right) + O\left(\frac{1}{L^6}\right), \quad (1)$$

where $m_\pi$ is the finite volume pion mass, $a_0$ is the scattering length for the isospin channel under consideration ($I = 0, 2$), and $c_1 = -2.837297$, $c_2 = 6.375183$ [1]. This result has been used in attempts to determine pion scattering lengths from numerical computations [2–4]. These computations, however, all used the quenched approximation, which raises the question as to how Lüscher's formula will change for quenched QCD. Here, we report on an analysis of this question [5] based on quenched chiral perturbation theory (ChPT) (for a review, see ref. [6]).

## 2. Definition of energy shift

In the numerical computations, the energy shifts $\Delta E \equiv E - 2m_\pi$ are extracted from euclidean correlation functions. For instance, for the $I = 2$ channel

$$\begin{aligned} C_{I=2}(t) &= \langle 0|\pi^+(t)\pi^+(t)\pi^-(0)\pi^-(0)|0\rangle_{\rm con} \quad (2)\\ &\equiv \sum_{|\alpha\rangle\neq|0\rangle} e^{-E_\alpha t}|\langle 0|\pi^-(0)\pi^-(0)|\alpha\rangle|^2 \\ &= Ze^{-2m_\pi t}\left(1 - \Delta E_{I=2}t + O(t^2)\right) + \ldots, \end{aligned}$$

where the dots indicate contributions from excited states. The fields $\pi^\pm(t)$ are zero spatial momentum fields: $\pi^\pm(t) = \sum_{\mathbf{x}} \pi^\pm(\mathbf{x}, t)$.

In the quenched approximation there is no justification for this parametrization in terms of energies and wavefunctions, since presumably no hamiltonian formalism exists. However, one may *define* $\Delta E$ directly from the euclidean correlation functions this way, as was done in refs. [2–4]. Note however that for instance the $O(t^2)$ terms will not follow the usual pattern [3]. Here we will follow this practice, and consider $\Delta E_{I=0,2}$ to one loop in euclidean quenched ChPT, with degenerate quark masses.

## 3. Role of the $\eta'$

The most drastic changes in eq. (1) will turn out to originate from the special role of the $\eta'$ in the quenched approximation. The quenched $\eta'$ twopoint function $D(p)$ contains a double pole

$$D(p) = \frac{1}{p^2 + m_\pi^2} - \frac{\mu^2}{(p^2 + m_\pi^2)^2} \quad (3)$$

($\mu^2$ is the parameter equivalent to the singlet part of the $\eta'$ mass in unquenched QCD) which leads to new infrared divergences in diagrams contributing to pion-pion scattering at one loop in ChPT. These diagrams contain the double pole in eq. (3) on one or both internal lines, as depicted in fig. 1 (for details, see refs. [5,6]).

*presented by MG

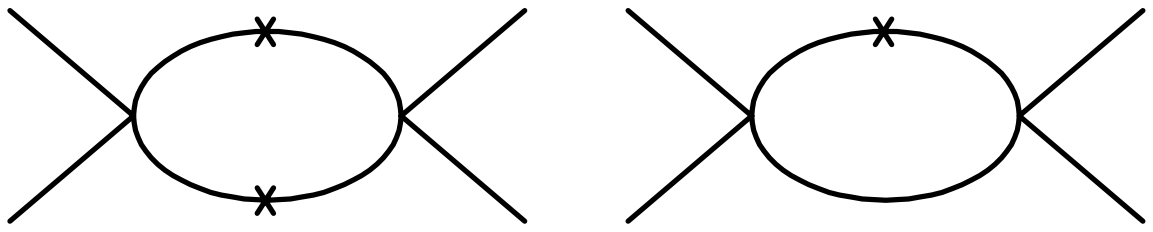

Figure 1. Examples of diagrams contributing to pion-pion scattering. The cross on the internal lines denotes a $\mu^2$ insertion (*cf.* second term in eq. (3)).

Defining

$$\delta \equiv \frac{\mu^2/3}{8\pi^2 f_\pi^2}, \qquad \epsilon \equiv \frac{m_\pi^2}{16\pi^2 f_\pi^2}, \tag{4}$$

where $f_\pi = 132\ MeV$ is the pion decay constant, the first diagram is $O(\delta^2)$ and the second $O(\delta\epsilon)$. $O(\delta^2)$ and $O(\delta\epsilon)$ contributions to pion-pion scattering do not have a counterpart in the unquenched theory, and are the ones that we will consider here (for a preliminary exploration in infinite volume, see ref. [7]).

## 4. Results

We will first give the results. Defining $\Delta E = \Delta E^{\rm tree} + \Delta E^{\rm 1-loop}$, we have

$$\Delta E^{\rm tree}_{I=0} = -\frac{7}{4 f_\pi^2 L^3}, \qquad \Delta E^{\rm tree}_{I=2} = \frac{1}{2 f_\pi^2 L^3} \tag{5}$$

and, for large $m_\pi L$

$$\begin{aligned} \frac{\Delta E^{\rm 1-loop}_{I=0}}{m_\pi} = & \\ & -0.319\delta^2 + 1.26\left(\frac{\delta^2}{8}+\delta\epsilon\right)\left(\frac{2\pi}{m_\pi L}\right)^2 \\ & - 0.159\left(\frac{5}{6}\delta^2+\delta\epsilon\right)\left(\frac{2\pi}{m_\pi L}\right)^3 \\ & + 0.127\delta^2\left(\frac{2\pi}{m_\pi L}\right)^4 \\ & - 0.0623\left(\delta^2 - \frac{128}{105}\delta\epsilon\right)\left(\frac{2\pi}{m_\pi L}\right)^6 + O(\epsilon^2), \end{aligned} \tag{6}$$

$$\frac{\Delta E^{\rm 1-loop}_{I=2}}{m_\pi} = -0.0531\left(\delta^2 - 6\delta\epsilon\right)\left(\frac{2\pi}{m_\pi L}\right)^3 + O(\epsilon^2). \tag{7}$$

Eqs (6,7) are valid up to corrections exponentially small in $m_\pi L$. On should in fact use the exact one-loop expressions for sizeable values of $m_\pi L$, for instance for $m_\pi L = 6$. Tables for a number of values of $m_\pi L$ can be found in ref. [5].

There are two important observations to be made from these results, in particular eq. (6). The first two terms on the righthand side show that "enhanced finite volume corrections" occur, which have no counterpart in the full theory (*cf.* eq. (1)). We will discuss the origin of this phenomenon below. The quenched result violates the pattern of eq. (1) in other ways, too. For instance, the coefficients of the $1/L^4$ terms are not proportional to $a_0^2$. We believe these are indications that the quenched theory is not well defined in Minkowski spacetime.

## 5. Origin of the enhanced finite volume corrections

Let us consider a correlation function like eq. (2), where two pions are created at time 0 and annihilated at time $t$. The simplest possible contribution is the one where the two pions do not interact at all, giving a contribution

$$\sim L^6 exp(-2m_\pi t) \tag{8}$$

(for this disconnected contribution there are two independent sums over the spatial volume). The second diagram in fig. 1 leads to a contribution

$$\begin{aligned} \sim \delta\epsilon L^3 e^{-2m_\pi t} & \\ \times \frac{1}{L^3}\sum_{\mathbf{k}} \int dt_1\, dt_2\, dt_\times\, e^{-2(\omega(\mathbf{k})-m_\pi)(t_1-t_2)}, \end{aligned} \tag{9}$$

where $\omega(\mathbf{k}) = \sqrt{m_\pi^2 + \mathbf{k}^2}$. The times $t_{1,2}$ correspond to the two four-pion vertices, and $t_\times$ to the $\mu^2$ insertion on the internal line. The enhanced terms come from that part of the integration region for which $0 < t_2 < t_\times < t_1 < t$. The integration over $t_\times$ is special to the quenched approximation, and comes from the double pole in eq. (3). It leads to an extra factor $t_1 - t_2$, and after integrating over $t_{1,2}$ we obtain a contribution linear in $t$

$$\sim \delta\epsilon \sum_{\mathbf{k}\neq 0} \frac{1}{(\omega(\mathbf{k}) - m_\pi)^2} t \tag{10}$$

(the $\mathbf{k}=0$ term in eq. (9) leads to an $O(t^2)$ contribution). For $\mathbf{k}=(2\pi/L,0,0)$ and $2\pi/m_\pi L \ll 1$ we have $\omega(\mathbf{k})-m_\pi \sim 1/L^2$, and therefore eq. (9) is of order $L^4$. Comparing with eq. (2) (using eq. (8) to determine $Z$) we finally get

$$\Delta E \sim \delta\epsilon/L^2, \qquad (11)$$

*i.e.* the $\delta\epsilon$ part of the second term on the right of eq. (6). A similar argument leads to a leading contribution $\Delta E \sim \delta^2$ coming from the first diagram in fig. 1. It is precisely the double pole term in the $\eta'$ twopoint function that leads to the enhanced finite volume corrections.

## 6. Excited states

Of course, apart from the lowest state contributions discussed sofar, there are also excited state contributions to eq. (2) of the form

$$\sim exp(-2\sqrt{m_\pi^2+\mathbf{k}^2}t), \quad \mathbf{k}\neq 0, \qquad (12)$$

which may contaminate the lowest state contribution in a numerical computation. The first excited state has $\mathbf{k}=(2\pi/L,0,0)$, and therefore, in order to have a good separation between the lowest state and the first excited state, one requires

$$\Omega \equiv 2m_\pi t\left(\sqrt{1+\left(\frac{2\pi}{m_\pi L}\right)}-1\right) \gg 1. \qquad (13)$$

However, in order to use eq. (2) to extract $\Delta E$, one also has to have that

$$|\Delta E^{\rm tree} t| \ll 1. \qquad (14)$$

Since $\Delta E$ is roughly inversely proportional to $L^3$ (eq. (5); the one-loop corrections have to be small for ChPT to be valid), it is nontrivial to satisfy both constraints simultaneously. To what extent they are satisfied (as well as whether ChPT is converging) has to be checked in each particular case of interest.

## 7. Numerical examples

We would like to illustrate our results with some numerical examples. We will first discuss a "real world" example, and then turn to the lattice computation of ref. [4].

### 7.1. *"Real world"*

We will take $m_\pi = 140\ MeV$, $f_\pi = 132\ MeV$, $\delta = 0.1$ [8] and $m_\pi L = 6$. One obtains [5]

| | $\Delta E^{\rm tree}$ | $\Delta E^{\rm 1-loop}$ |
|---|---|---|
| $I=0$ | $-1.3\ MeV$ | $-0.3\ MeV$ |
| $I=2$ | $0.36\ MeV$ | $-0.07\ MeV$ |

(15)

For this example, the one-loop corrections are of order 25% of the tree-level terms. For larger volumes the one-loop corrections become large relative to the tree-level terms. For instance, for $m_\pi L = 8$, $\Delta E^{\rm 1-loop}/\Delta E^{\rm tree}|_{I=0} = 63\%$ and $\Delta E^{\rm 1-loop}/\Delta E^{\rm tree}|_{I=0} = 240\%$ for $m_\pi L = 12$. Clearly, quenched ChPT breaks down for these values of $L$. This is due to the enhanced finite volume terms in eq. (6). And, as one would expect, ChPT also breaks down for small values of $m_\pi L$. ($m_\pi L = 4$ is still ok.) We should note that a larger value for $\delta$ also increases the one-loop corrections.

To conclude this example, we note that if we choose $t$ such that $|\Delta E^{\rm tree} t| \simeq 0.1$, we have $\Omega_{I=0} = 10$ and $\Omega_{I=2} = 35$.

### 7.2. *Computation of Kuramashi et al.*

We now turn to the most recent lattice determination of pion scattering lengths. Kuramashi *et al.* [4] performed a lattice computation on a $12^3\times 20$ lattice at $\beta = 5.7$. They used Wilson fermions with $af_\pi = 0.143$ and $am_\pi = 0.508$, and staggered fermions with $af_\pi = 0.187$ and $am_\pi = 0.29$. They did not include the so-called "double-annihilation" diagram, which leads to a somewhat different expression for $\Delta E^{\rm 1-loop}_{I=0}$ [5]. Again with $\delta = 0.1$, we obtain for their staggered fermion parameters

| | $a\Delta E^{\rm tree}$ | $a\Delta E^{\rm 1-loop}$ |
|---|---|---|
| $I=0$ | $-0.029$ | $0.0002$ |
| $I=2$ | $0.0083$ | $-0.0017$ |

(16)

and for their Wilson fermion parameters

| | $a\Delta E^{\rm tree}$ | $a\Delta E^{\rm 1-loop}$ |
|---|---|---|
| $I=0$ | $-0.050$ | $0.005$ |
| $I=2$ | $0.014$ | $0.001$ |

(17)

The one-loop corrections in this case are less than 20%.

Kuramashi *et al.* used a fitting range $4 \leq t \leq 9$, which leads to the following values for the quantities of eqs. (13,14):

$$staggered: \quad |\Delta E^{\rm tree}_{I=0} t| = 0.26 \quad @ \quad t = 9, \qquad (18)$$
$$\Omega_{I=0} = 2.4 \quad @ \quad t = 4.$$

$$Wilson: \quad |\Delta E^{\rm tree}_{I=0} t| = 0.45 \quad @ \quad t = 9, \qquad (19)$$
$$\Omega_{I=0} = 1.8 \quad @ \quad t = 4,$$

Clearly, one would like to have a considerably larger separation of the six excited states with $|\mathbf{k}| = 2\pi/L$.

## 8. Conclusion

We have calculated the two-pion finite volume energy shifts in the quenched approximation, using quenched ChPT. The enhanced finite volume corrections that show up at one loop in the energy shifts (and do not have a counterpart in the unquenched theory) are yet another example of the bad infrared behavior of the quenched approximation. These corrections lead to a breakdown of ChPT in *large* volumes. With examples we also showed that reasonable choices of parameters exist for which the one-loop corrections are moderately small relative to the tree-level contributions.

It should be clear from our discussion that the quenched one-loop terms have no relation to those in the full theory. In order for the quenched theory to be close to the unquenched theory, a minimal requirement therefore is that one-loop contributions in *both* theories be small. For pion scattering lengths this implies that the minimal error from quenching is of order 25%, since that is the size of one-loop corrections in unquenched ChPT [9].

Finally, we would like to note that apparently only the euclidean quenched theory is well defined. A nonperturbative, hamiltonian analysis along the lines of the second reference of [1] does not seem possible. Nevertheless, one may formally define quenched pion scattering lengths from the $1/L^3$ terms in $\Delta E$ [5]. One finds that they are linearly divergent in the pion mass at one loop, again as a consequence of the peculiar role of the $\eta'$ in the quenched theory.

**Acknowledgements**

We would like to thank Steve Sharpe, Akira Ukawa and Pierre van Baal for useful discussions. CB and MG are supported in part by the Department of Energy under grant #DOE-2FG02-91ER40628, and MG by a DOE Outstanding Junior Investigator grant.

## REFERENCES

1. M. Lüscher, Comm. Math. Phys. 105 (1986) 153; Nucl. Phys. B354 (1991) 531.
2. M. Guagnelli, E. Marinari and G. Parisi, Phys. Lett. B240 (1990) 188.
3. S.R. Sharpe, R. Gupta and G.W. Kilcup, Nucl. Phys. B383 (1992) 309; R. Gupta, A. Patel and S.R. Sharpe, Phys. Rev. D48 (1993) 388.
4. Y. Kuramashi *et al.*, Phys. Rev. Lett. 71 (1993) 2387; Nucl. Phys. B (Proc. Suppl.) 34 (1994) 117; hep-lat/9501024.
5. C.W. Bernard and M.F.L. Golterman, hep-lat/9507004.
6. M.F.L. Golterman, Acta Phys. Pol. B12 (1994) 1731 and refs. therein.
7. C.W. Bernard *et al.*, Nucl. Phys. B (Proc. Suppl.) 34 (1994) 334.
8. R. Gupta, Nucl. Phys. B (Proc. Suppl.) 42 (1995) 85.
9. J. Gasser, to be published in the 2nd edition of the DAΦNE Physics Handbook, eds. L. Maiani, G. Pancheri and N. Paver, hep-ph/9412392.